\documentclass[9pt, conference]{IEEEtran}

\usepackage{amsmath,graphicx}
\usepackage{multirow}
\usepackage{hhline}

\usepackage{footnote}
\usepackage{epsfig}
\usepackage{latexsym}
\usepackage{psfrag}
\usepackage{pstricks}
\usepackage{dsfont}
\usepackage{cancel}
\usepackage{epstopdf}
\usepackage{bigints}
\usepackage{cite}
\usepackage{url}
\usepackage[normalem]{ulem}
\usepackage{multirow}

\usepackage{amsfonts}
\usepackage{amssymb}
\usepackage{spconf}

\begin{document}

\title{CNN-based detection of generic contrast adjustment with JPEG post-processing}
%
\name{M.Barni$^{\#}$, A.Costanzo$^{\dagger}$, E.Nowroozi$^{\#}$,  B.Tondi$^{\#}$
\thanks{This work has been partially supported by a research sponsored by DARPA and Air Force Research Laboratory (AFRL) under agreement number FA8750-16-2-0173. The U.S. Government is authorised to reproduce and distribute reprints for Governmental purposes notwithstanding any copyright notation thereon. The views and conclusions contained herein are those of the authors and should not be interpreted as necessarily representing the official policies or endorsements, either expressed or implied, of DARPA and Air Force Research Laboratory (AFRL) or the U.S. Government.\newline * The list of authors is provided in alphabetic order.}
\vspace{-0.2cm}
}
\address{$^{\#}$ Department of Information Engineering and Mathematics, University of Siena\\
$^{\dagger}$ CNIT - Consorzio Nazionale Interuniversitario per le Telecomunicazioni}

%
%

%

\maketitle
\begin{abstract}
Detection of contrast adjustments in the presence of JPEG post processing is known to be a challenging task. JPEG post processing is often applied innocently, as JPEG is the most common image format, or it may correspond to a laundering attack, when it is
purposely applied  to erase the traces of manipulation.
In this paper, we propose a CNN-based detector for generic contrast adjustment, which is robust to JPEG compression.  The proposed system relies on a patch-based Convolutional Neural Network (CNN), trained to distinguish  pristine images from contrast adjusted images, for some selected adjustment operators of different nature.
Robustness to JPEG compression is achieved
by training
the CNN with JPEG examples, compressed
over a range of Quality Factors (QFs). Experimental results show that the detector works very well
and scales well with respect to the adjustment type, yielding very good performance under a large variety of unseen tonal adjustments.
\end{abstract}

\section{INTRODUCTION}
\label{sec:intro}

\enlargethispage{\baselineskip}

Adjustment of contrast and lighting conditions of image subparts is often performed during forgery creation. Therefore, the problem of detecting such manipulation has been widely studied in image forensics, and, more recently, in scenarios encompassing the presence of an adversary \cite{Boh12,BarGon13}.
Due to the peculiar traces left by contrast adjustment operators, most early works were based on the analysis of first order statistics \cite{stamm2010forensic, cao2010forensic, cao2014contrast}.
Such approaches, however, are easily circumvented by the adversary, by means of both targeted \cite{cao2010anti} and also universal approaches \cite{barni2012universal}. To cope with such attacks, countermeasures were developed in turn, based on a second-order analysis \cite{pan2011exposing,de2015second}.
%
%
However, in most cases, the attack is of laundering-type, consisting in the application of a post-processing operation, e.g., a geometric transformation, filtering or compression. Laundering attacks have been shown to be very powerful against manipulation detectors in general \cite{li2016identification}.
In particular, the performance of contrast manipulation detectors proposed so far decrease significantly in the presence of even mild post-processing and, above all, they all exhibit  poor robustness against JPEG compression \cite{stamm2010forensic,cao2014contrast,pan2011exposing,li2016identification,singh2016analysis}, even when the compression is very weak.
Since images are often stored and distributed in JPEG format, JPEG compression is also one of the most common post-processing images are subject to. Therefore, designing a JPEG-robust contrast adjustment detector is of primary importance.

%
%
%
In this paper, we face with the above problem by resorting to JPEG-aware data-driven classification \cite{barni2017eusipco}, that is, by designing a data driven detector for contrast adjustment which is trained to recognise the specific class of JPEG laundering attacks. In particular, we look for a generic detector of contrast adjustment, that is, a detector which generalizes well to a wide variety of tonal adjustments.
The proposed method relies on a Convolutional Neural Network (CNN) architecture.
The CNN is directly fed with image pixels (with no pre-processing), hence the discriminative features for our problem are self-learned by the CNN.
%
%
%
%
Specifically, the proposed detector relies on a JPEG-aware, patch-based CNN, which is used to classify image regions, i.e. image patches. A test image is then divided into patches which are tested separately by feeding them to the CNN. The soft patch scores (CNN outputs) are collected and the global decision on the image is performed on the score vector.
%
All the compression QFs inside a range of values are considered to train the CNN.
Noticeably,
we could also exploit the knowledge of the QF, which can be estimated from the image header, and specialize the CNN to work for one QF only (hence training several CNNs). However, such an approach has the drawback
of being easily prone to attacks: just re-saving the image in uncompressed format (e.g., PNG,..) or compressing again the image with a different QF would prevent the correct identification of the QF used to compress the image. 
%
Therefore, for our global manipulation detection task, we considered only one CNN model; the final detection accuracy is raised by exploiting the fact that patches coming from a same image are generated under the same hypothesis (being all pristine or contrast adjusted patches), and hence should all result in a small (or large) soft value as CNN output.

Experiments show that our system achieves good performance over a wide range of QFs. Thanks to the fact that  the CNN is simultaneously trained with different contrast adjustments, our detector
achieves good scalability with respect to the contrast adjustment type, yielding good performance  over a large variety of contrast, brightness and tonal adjustments, i.e. under processing mismatch conditions.
Good performance are maintained in the absence of JPEG, that is, when the contrast adjustment is the last step of the manipulation chain.

The rest of the paper is organized as follows: in Section \ref{sec.problemForm}, we define the detection task we focus on, describe the proposed CNN-based detector and the network architecture. In Section \ref{sec.expMethod}, we first detail the methodology followed for conducting our experiments, then we report and discuss the results.


\enlargethispage{\baselineskip}

\section{PROPOSED SYSTEM}
\label{sec.problemForm}

\begin{figure}
\centering
\includegraphics[width=0.8\columnwidth]{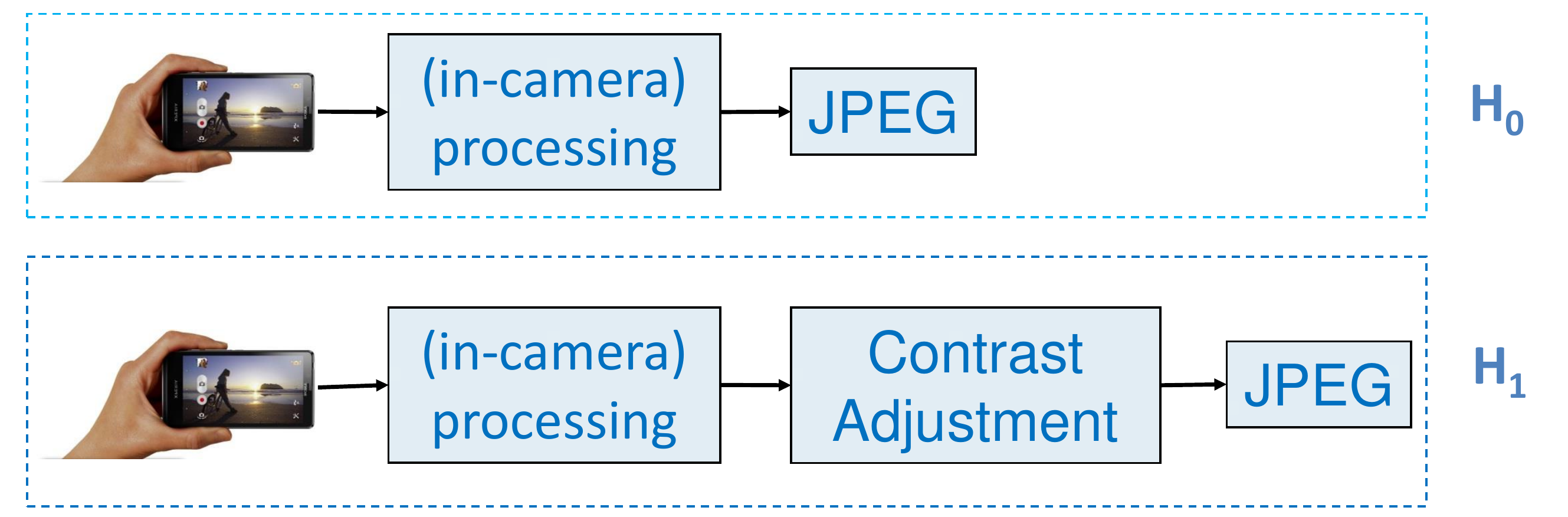}
\caption{Detection task considered in this paper.} 
\label{Scheme}
\end{figure}

Figure \ref{Scheme} schematizes the problem addressed in this paper.
We let hypothesis $H_0$ correspond to the case of pristine images and $H_1$ to the case of contrast adjusted  images In both cases, the image is JPEG compressed at the end, with a given QF. 
In this scheme, JPEG compression can also be viewed as a counter-forensic, laundering-type, attack, due to its known effectiveness in erasing the traces of contrast manipulations \cite{stamm2010forensic,cao2014contrast,pan2011exposing,li2016identification,singh2016analysis}.
%

\begin{figure}
\centering
\includegraphics[width=1\columnwidth]{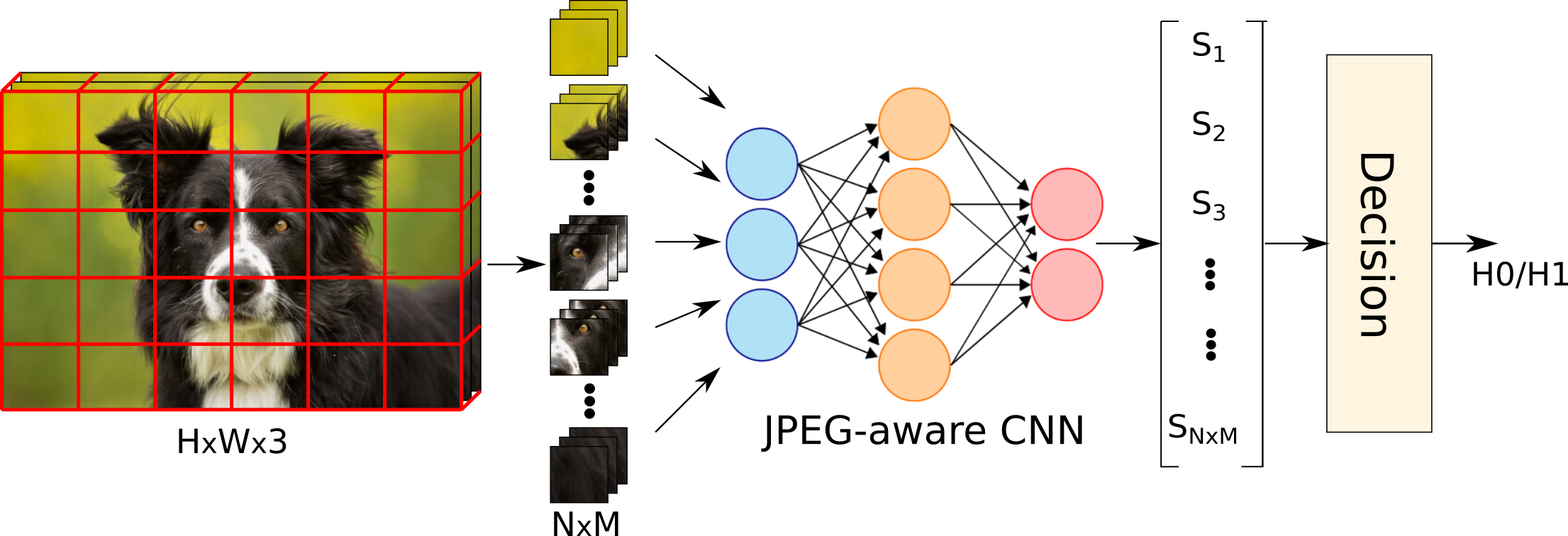}
\caption{Pipeline of the proposed generic, JPEG-aware, contrast adjustment detector. Adaptive histogram equalization, gamma correction (both compression and expansion) and histogram stretching are used to train the network.}
\label{Pipeline}
\end{figure}

%
%
%
The architecture of the proposed detection scheme is reported in Figure \ref{Pipeline}.
The color image is divided into non-overlapping patches of size $64\times 64$ which are fed to a JPEG-aware CNN detector.
The patch scores, i.e. the CNN soft outputs for all the patches, are then collected and the final decision is based on the score vector $s = (s_1, s_2,...s_{N\times M})$ (where $N\times M$ is the total number of blocks). The decision is made by simply thresholding the sum of the scores, i.e. according to the statistic\footnote{This is a simple and non optimized choice. Other possible fusion strategies could be adopted.}  $\sum_i s_i/(M\cdot N)$. Since patches coming from the same image are drawn under the same hypothesis, such normalised sum is expected to be large in one case (contrast adjusted image) and small in the other (pristine image).

The JPEG-aware CNN is trained with JPEG compressed images on one hand ($H_0$) and images subject to \ contrast adjustment followed by JPEG compression on the other ($H_1$). The network architecture and the training strategy are detailed in the following sections.
In the attempt to build a detector which generalizes to unseen adjustments, we consider contrast adjustments of different nature to train the network. Specifically, the  processing we selected are: adaptive histogram equalization, gamma correction (both compression and expansion of the contrast) and histogram stretching.

Regarding the compression QF, we focus on values ranging from medium-high to high values (i.e., $QF \ge 80$), which are commonly used in many practical applications.

\enlargethispage{\baselineskip}

\subsection{CNN architecture}

Our first attempts to train a network for our problem by using architectures similar to those adopted for other forensic tasks \cite{ChenMFwithCNN,StammCNNuniv,barni2017aligned} were unsuccessful.
A possible explanation is the following:
while some processing operations, e.g. local filtering and double JPEG, introduce  local patterns that a properly trained CNN with few layers is able to 'easily' learn, common contrast adjustments do not leave local visual artifacts, thus making self CNN learning harder and calling for the adoption of deeper models.
We were in fact able to get higher accuracies by switching to deeper architectures, with small kernel sizes and small strides of the convolutional layers, inspired by those adopted in image classification and pattern recognition applications \cite{simonyan2014very}. In particular, as suggested in \cite{simonyan2014very}, we adopt a kernel size of $3\times 3$ and stride  1 for all the convolutional layers, and only 1 fully connected layer. We set the number of convolutional layers to 9, which, although lower than that adopted in  \cite{simonyan2014very} (16-19), is still a significant depth compared to those commonly considered for forensic tasks \cite{ChenMFwithCNN,StammCNNuniv,barni2017aligned} (up to 4-5).

\begin{figure*}[ht]
\centering
\includegraphics[width=0.9\textwidth]{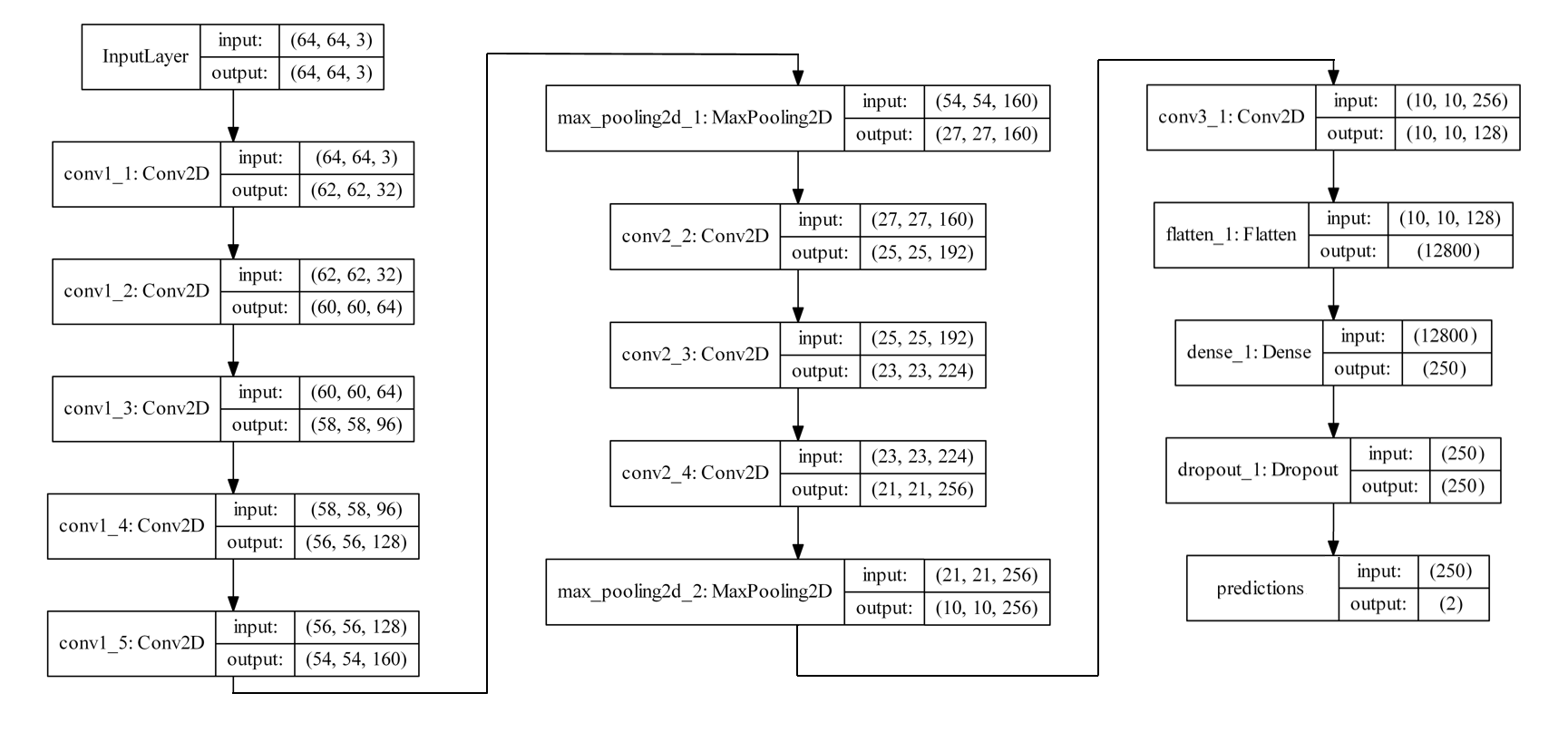}
\vspace{-0.5cm}
\caption{Architecture of the proposed network.}
\label{Net_scheme}
\end{figure*}

More specifically, the structure of our network for patch classification (see Figure \ref{Net_scheme}) is detailed as follows: it takes a color patch of size $64 \times 64$ as input and consists of
\begin{itemize}
\item {\bf 5 convolutional layers}
     followed by a {\bf max-pooling} layer. In the first convolutional layer 32 filters are applied. Then, the number of filters increases by 32 at each layer. For all the filters, the kernel size is $3 \times 3$ and the stride is always 1. Max-pooling is applied with kernel size $2 \times 2$ and stride $2$ producing a final $27\times 27\times 160$ feature map.
\item {\bf 3 convolutional layers} followed by a {\bf max-pooling} layer.  As before, the number of filters of size $3\times 3$ (applied with a stride 1) increases by 32 at each layer. The pooling is the same as before, yielding a $10 \times 10 \times 256$ feature map.
    \item A final {\bf convolutional layer} with 128 filters of size $1\times 1$ generating a  $10 \times 10 \times 128$ feature map.
    \item A {\bf fully-connected layer} with 250 input neurons, dropout 0.5, and 2 output neurons, followed by a softmax layer (last 3 blocks in the scheme of Figure \ref{Net_scheme}).
    \end{itemize}
Some comments regarding the main features of the above architecture are in order: the use of many convolutions (5) before the first pooling layer permits to consider a large receptive field for each neuron, which is good to capture relationships among pixels in large neighborhoods; the stride 1 permits to retain as much spatial information as possible.
The purpose of the final convolutional layer is to reduce the number of parameters by halving the number of maps (from 256 to 128), without affecting spatial information.
The adoption of only one fully connected layer also permits to reduce the number of parameters without affecting too much the performance.
Finally, we observe that using small patches ($64\times 64$) permitted us to increase the depth of the network for the same number of parameters. The use of small patches is also suitable for tampering localization (the detection accuracy is then raised by aggregating the patch scores).

\subsection{CNN training strategy}

We obtained the JPEG-aware CNN model by training the network in two steps. First, the network is trained to recognize between patches coming from pristine and contrast-adjusted images for the uncompressed case, getting an (unaware) pre-trained model.  Then, the aware model is obtained by fine-tuning the unaware network, by feeding the CNN with JPEG compressed examples of the above classes.
Since the network is pretty deep and then the number of images used for training is very large,
we performed compression on-the-fly by augmenting the data inside the network; hence, the compression is performed directly on the $64\times 64$ patches (that is, after image splitting).
Such a strategy is viable because the JPEG compression is a local operator which
can be applied separately on multiples of $8\times8$ image patches producing the same result as if it were applied on the entire image.

\enlargethispage{\baselineskip}

\section{EXPERIMENTS}

\subsection{Methodology}
\label{sec.expMethod}

We built the training and testing sets by starting from color images in uncompressed format. The images for the $H_0$ and $H_1$ classes were produced as detailed in Figure \ref{Scheme}.
The adjustment of the contrast under $H_1$ is obtained by considering several algorithms. As we said, to generate the images used for training, we considered the following operators: Adaptive Histogram Equalization (in particular, its refined, Contrast Limited, implementation, \verb"CLAHE" \cite{Zuiderveld1994}), Gamma Correction (\texttt{$\gamma$ Corr}), and Histogram Stretching (\verb"HS").
Such operators are designed for one-channel images; to make them work on color images, we applied them as follows: for the images processed with \verb"CLAHE",
we first converted the images from  RGB to HSV, we applied the enhancement to the luminance channel only, namely the V channel, and converted them back to the RGB domain
\footnote{The straightforward application of  \texttt{CLAHE} (and \texttt{HS}) to each channel separately unnaturally changes the color balance and produces visually unpleasant images.}.
The same strategy is adopted to generate the images processed with \verb"HS". Finally, for the \texttt{$\gamma$ Corr}, the contrast is modified by applying the operator to each channel (R, G and B) separately. The above operators are applied in equal percentage to generate the class of contrast adjusted images ($H_1$).
Regarding the parameters,
the clip-limit parameter for  \verb"CLAHE" is set to 0.005, the $\gamma$ value  to 1.5 and 0.7 (randomly chosen with probability 0.5), and the saturation percentage of the \verb"HS" to 5\% for both black and white values. The above choices do not introduce visually unpleasant artifacts.

For generating the test images,  we also considered different values of the parameters for the same operators (to assess the performance under parameter mismatch), and different operators, by processing the images with adjustment tools provided by Photoshop.
In particular, we considered the following tonal adjustments:
\begin{itemize}
\item \verb"AutoContrast", \verb" AutoColor" and  \verb"AutoTone";
   algorithms which operate differently with respect to the color channels. The clipping is set to 7\% for \verb"AutoContrast" and \verb"AutoColor"  and to $5\%$ for \verb"AutoTone"; the snap neutral midtones option is selected for the \verb" AutoColor";
 \item  \verb"Curves_S";
  a (hand-made) smooth S-curve is applied to enhance the contrast in the midtones;
  \item \verb"Brightness" and  \verb"Contrast";
  generic tools for enhancing and reducing brightness and contrast; for the enhancement, we set \verb"Brightness" to 50 (\verb"Brightness+") and \verb"Contrast" to 70 (\verb"Contrast+"), while for the reduction, we set  \verb"Brightness" to -70 ( \verb"Brightness-") and \verb"Contrast" to -50 (\verb"Contrast-");
 \item Histogram Equalization (\verb"HistEq").
 \end{itemize}
The  \verb"HistEq" manipulation is considered for completeness:
although its visual impact is much stronger with respect to that of the other manipulations, and hence is rarely adopted in practice,  the \verb"HistEq" manipulation is often considered in multimedia forensic literature.

Regarding JPEG compression, we randomly selected the $QF$s (uniformly) in the range $[90,100]$ to compress the images used for training. For testing, we also considered images compressed with QF $85$ and $80$.
%
%
%

%

\subsection{Results}
\label{sec.expRes}

Uncompressed, camera-native, images (.tiff)  are taken from the RAISE8K dataset \cite{RAISE8K} (of size $4288 \times 2848$), splitted into training and test set, and then contrast-adjusted to produce the images for $H_1$ in the unaware case (i.e., without the final JPEG).
The images are then divided into $64 \times 64$ patches for CNN training and testing:  $2\times 10^{6}$ patches per class (coming from more than 1000 training images) were selected to train the CNN, whereas $2\times 10^{5}$ patches were used for testing. In the aware case, the patches are JPEG compressed with QF $\in [90,100]$.
The overall performance of the detector are tested on 300 images from the test set, both uncompressed and compressed with QF = $ \{100, 98,95,90,85,80\}$.
The images used for training were all processed with the OpenCV library for Python.
For the tests, the Photoshop software was also adopted.
We used the TensorFlow framework, via the Keras API \cite{Keras}, to implement our CNN.
We ran our experiments using 2x Asus GeForce GTX1080TI - 11GB DDR5 gpu.
The Adam solver is used with learning rate $1e-4$ and momentum $0.99$.
We set the batch size for training and testing to 32 images. Both the aware and unaware models are trained on 3 epochs.

When training and testing are performed with uncompressed images (unaware case),
the average test accuracy of the CNN on image patches is $93.5\%$,
where the average is taken on the 3 manipulations, i.e., \verb"CLAHE", \texttt{$\gamma$ Corr} (compression and expansion) and \verb"HS", and on all the QFs inside the training range. For the overall system,
we get almost perfect classification, that is, the Area Under Curve (AUC) is $99,8\%$, which is in line with the state of the art \cite{li2016identification}. A noticeable strength is that here these performance are achieved by one (generic) system only, rather than using separate systems each one specialized on one manipulation.
By testing the unaware detector with JPEG compressed images, the performance drop to AUC = $56\%$
thus showing that
the CNN model is not robust to the JPEG laundering attack.

\enlargethispage{\baselineskip}

Concerning the aware case, the average accuracies that we obtained at the patch level in the range of QFs  $[90,100]$ are: $0.84$ for \verb"CLAHE", $0.72$ for \texttt{$\gamma$ Corr} and $0.79$ for \verb"HS".
{These accuracies are not very high; however, the performance are moderately good with respect to all the contrast adjustment operators.
%
We also observe that specializing our network to work with one QF only, we could have obtained higher performance at the patch level; however, as said before, to be robust against common manipulations (as recompression and saving in uncompressed formats), we look for a detector of generic contrast adjustments which works well on a range of $QF$s.
\begin{table}
\setlength{\tabcolsep}{4pt}
\begin{center}
\caption{Performance (AUC) of the detector under matched processing. The matched parameters are in bold.}
\vspace{0.2cm}
\begin{tabular}{  c  | c | c  c  c c  c c c c }
\multicolumn{2}{c}{} & \multicolumn{8}{c}{\bf QF}   \\
\multicolumn{2}{c|}{} &  \small{{\bf no jpg}} & {\bf 100} & {\bf 98} & {\bf 95} & {\bf 90}  & {\bf 85} & {\bf 80} & {\bf 75} \\ \hline
    \multirow{3}{*}{{\bf CLAHE}} & 0.003 & 100 & 99.9  & 99.8 & 98.9 & 97.6 & 97.1 & 96.8 & 96\\
    & {\bf 0.005}  & 100 & 99.9  &  99.9 &  99.4 & 98.9 & 98.8   & 98.5 & 98 \\
    & 0.007 & 100 & 99.9  & 100 & 99.6 &  99.1  & 98.9 & 98.7 & 98.5\\
    \hline
    \multirow{4}{*}{\bf $\gamma$ Corr}
    &  {\bf 1.5}  & 98.8 & 98.5  & 94.2 & 89.2 &  87& 84  & 81.2  & 81 \\
    & 1.7 & 99.4  &  98.9 & 95.7 & 91.8 & 90.4  & 89.7  &  89.2 & 88.1 \\
    & {\bf 0.7}  & 99.1 & 97.1  & 92.3 & 87.3 &  85.6 & 81  & 78 & 69 \\
    & 0.6  &  99.7 & 99.5   & 97.3 & 91.6 &  86.7 & 83.7  & 80.1 & 77.3 \\ \hline
    \multirow{3}{*}{{\bf HS} (\%)} & 3 & 99.6 & 98.1 & 95.8 & 91.4  & 87.8  &  85.7 & 83.5 & 83 \\
    & {\bf 5} & 99.5 & 98.9  & 97.6 & 93.7 & 92.6 & 91.5 &  90.3 & 89.4\\
    & 7 & 100 & 99.3 & 98.3 & 95.5 & 94  & 93.7  & 93.6 & 93\\
    \hline
\end{tabular}
 \label{tab.aware_fullImage}
\end{center}
\vspace{-0.5cm}
\end{table}
%
%
The overall performance of the detector on full images are reported in Table \ref{tab.aware_fullImage} in terms of AUC, for both matched and mismatched processing parameters.
%
%
%
The \texttt{CLAHE} manipulation is the easiest to detect (the AUC is always above $98\%$). The most difficult case corresponds to \texttt{$\gamma$ Corr}, where the AUC is below $90\%$ for QF $\le 95$. This behavior is due to the fact that such kind of adjustment is difficult to detect by itself and above all to the fact that the CNN is simultaneously trained with values smaller and larger than 1, corresponding to a compression and an expansion of the contrast.\footnote{We verified that if the detector is trained with $\gamma = 1.5$  only (gamma expansion), the AUC for the $\gamma$ Corr is above $97\%$ for every QF $\ge 85$. In this case, however, the performance with respect to a compression ($\gamma < 1$) are very poor even with large QF (e.g. AUC $= 78\%$ for QF $= 95$).}
These results significantly improve those achieved by the SVM-based approaches from the literature, where, depending on the specific contrast adjustment considered, the AUC may drop to about 60\% \cite{singh2016analysis}, and 72\% \cite{li2016identification} (on the average) in the presence of JPEG compression in the same range.

We observe that the performance are good in the presence of a mismatch in the processing parameters: better performance are obtained when the adjustment is stronger than in the matched case, and worse when it is weaker.
The performance remain very good in the absence of JPEG: the AUC is $99.6\%$ on the average in the matched case, which is in line with the AUC achieved by the unaware detector. Expectedly, performance decrease as QF decreases.
However, good robustness to JPEG compression is achieved (at least for \verb"CLAHE" and \verb"HS") also when the QF is $85$ and $80$, which are outside the training range, whereas, below 80, performance  become poorer.
It is worth observing that, for a fixed false alarm rate, the threshold on the aggregated score changes by varying the $QF$: specifically, for a  false alarm of $5\%$, the threshold ranges in $[0.56 : 0.71]$.
Note that, since the last compression $QF$ is always known (or it can be estimated), such a variability of the threshold is not a problem.
%
Table \ref{tab.aware_fullImage_PS} shows the results under various contrast/brightness adjustment performed with Photoshop.
Based on these results, we can argue that the CNN-based detector scales well with respect to the adjustment type maintaining good performance when the tones of the image are adjusted in different ways and, possibly, selectively in different tonal ranges (\verb"Curve_S"), and when the adjustment operates differently on the color channels (the \verb"Auto" processing).
The AUC is large with respect to all the $QF$s for some of the processing (\verb"AutoTone", \verb"Curve_S", \verb"HistEq") and, in general, it remains above $90\%$ in most of the cases.


\begin{table}
\setlength{\tabcolsep}{4pt}
\begin{center}
\caption{Performance (AUC) of the detector for different tonal adjustments.}
\begin{tabular}{  c | c  c  c c  c c c  }
\multicolumn{1}{c}{} & \multicolumn{7}{c}{{\bf QF}}   \\
& \small{\bf no jpg} & {\bf 100} & {\bf 98} & {\bf 95} & {\bf 90}  & {\bf 85} & {\bf 80} \\ \hline
\small{\verb"HistEq"} & 100 & 99.9  & 99.9 & 99.5 &  98.3 & 96.9  & 94.8 \\ \hline
\small{\verb"Brightness+"}  & 97.5 & 97.7 & 95.2 & 93.6 & 91.2 & 87.8 & 85.6\\ \hline
\small{\verb"Contrast+"}  & 99.1 & 100  & 99.6 & 97.9 & 94.7 & 91.9 & 87.1 \\ \hline
\small{\verb"Brightness-"} & 96.7 & 97.3 & 93.3 & 90.1 & 84.2 & 78.8 & 75.6\\ \hline
\small{\verb"Contrast-"}  &  98.8 & 99.6 & 96.4 & 91.2 & 87 &  82 & 80 \\ \hline
\small{\verb"Curve_S"}  & 99.6 & 99.8  & 99.8 & 99.1 & 97.7 & 96 & 93.6\\ \hline
\small{\verb"AutoContrast"}  & 95.9 & 94.7 & 93 & 91.9 & 90.2 & 89 & 86.5\\ \hline
\small{\verb"AutoColor"}  & 98.2 & 98.6 & 96.8 & 95.3 & 93.7 & 91.8 & 89.1 \\ \hline
\small{\verb"AutoTone"}  & 99.5 & 99.5 & 99 & 98.2 & 97.2 & 96.1 & 94.5\\
\hline
\end{tabular}
\label{tab.aware_fullImage_PS}
\vspace{-0.7cm}
\end{center}
\end{table}

\section{CONCLUSIONS} 
\label{sec.con}

We proposed a JPEG-aware CNN-based approach to cope with the well known problem of detection of contrast adjusted images in the presence of JPEG post-processing.
To accomplish this task,
and build a detector which works well for generic contrast adjustment, we trained the CNN with a certain number of adjustments of different nature. Results show that our detector achieves good performance over a wide range of QFs and generalizes well to unseen tonal adjustments. As further research, it would be interesting to see if the performance with respect to the most difficult cases can be improved by refining the composition of the training, i.e., the types of contrast adjustments considered and their proportions,
and also the fusion strategy at the final stage.
As a future work, we would like to improve the performance at the  patch level to move from detection to localization.

\section{ACKNOWLEDGMENTS} 
This work has been partially supported by DARPA and AFRL under the research grant number FA8750-16-2-0173. The United States Government is certified to reproduce and distribute reprints for Governmental objectives notwithstanding any copyright notation thereon. The views and conclusions consist of herein are those of the authors and should not be interpreted as necessarily representing the official policies or endorsements, either expressed or implied, of DARPA and AFRL or U.S. Government.


\bibliographystyle{IEEEbib}
\bibliography{ICIP18}

\begin{thebibliography}{10}

\bibitem{Boh12}
Rainer {B\"{o}hme} and Matthias Kirchner,
\newblock ``Counter-forensics: Attacking image forensics,''
\newblock in {\em Digital Image Forensics}, H.~T. Sencar and N.~Memon, Eds.
  Springer Berlin / Heidelberg, 2012.

\bibitem{BarGon13}
Mauro Barni and Fernando {P{\'e}rez-Gonz{\'a}lez},
\newblock ``Coping with the enemy: advances in adversary-aware signal
  processing,''
\newblock in {\em ICASSP 2013, IEEE International Conference on Acoustics,
  Speech and Signal Processing}, Vancouver, Canada, 26-31 May 2013, pp.
  8682--8686.

\bibitem{stamm2010forensic}
Matthew~C Stamm and KJ~Ray Liu,
\newblock ``Forensic detection of image manipulation using statistical
  intrinsic fingerprints,''
\newblock {\em IEEE Transactions on Information Forensics and Security}, vol.
  5, no. 3, pp. 492--506, 2010.

\bibitem{cao2010forensic}
Gang Cao, Yao Zhao, and Rongrong Ni,
\newblock ``Forensic estimation of gamma correction in digital images,''
\newblock in {\em 2010 17th IEEE International Conference on Image Processing
  (ICIP)}. IEEE, 2010, pp. 2097--2100.

\bibitem{cao2014contrast}
Gang Cao, Yao Zhao, Rongrong Ni, and Xuelong Li,
\newblock ``Contrast enhancement-based forensics in digital images,''
\newblock {\em IEEE Transactions on Information Forensics and Security}, vol.
  9, no. 3, pp. 515--525, 2014.

\bibitem{cao2010anti}
Gang Cao, Yao Zhao, Rongrong Ni, and Huawei Tian,
\newblock ``Anti-forensics of contrast enhancement in digital images,''
\newblock in {\em Proceedings of the 12th ACM Workshop on Multimedia and
  Security}. ACM, 2010, pp. 25--34.

\bibitem{barni2012universal}
Mauro Barni, Marco Fontani, and Benedetta Tondi,
\newblock ``A universal technique to hide traces of histogram-based image
  manipulations,''
\newblock in {\em Proceedings of the on Multimedia and security}. ACM, 2012,
  pp. 97--104.

\bibitem{pan2011exposing}
Xunyu Pan, Xing Zhang, and Siwei Lyu,
\newblock ``Exposing image forgery with blind noise estimation,''
\newblock in {\em Proceedings of the thirteenth ACM multimedia workshop on
  Multimedia and security}. ACM, 2011, pp. 15--20.

\bibitem{de2015second}
Alessia De~Rosa, Marco Fontani, Matteo Massai, Alessandro Piva, and Mauro
  Barni,
\newblock ``Second-order statistics analysis to cope with contrast enhancement
  counter-forensics,''
\newblock {\em IEEE Signal Processing Letters}, vol. 22, no. 8, pp. 1132--1136,
  2015.

\bibitem{li2016identification}
Haodong Li, Weiqi Luo, Xiaoqing Qiu, and Jiwu Huang,
\newblock ``Identification of various image operations using residual-based
  features,''
\newblock {\em IEEE Transactions on Circuits and Systems for Video Technology},
  2016.

\bibitem{singh2016analysis}
Neetu Singh and Abhinav Gupta,
\newblock ``Analysis of contrast enhancement forensics in compressed and
  uncompressed images,''
\newblock in {\em 2016 International Conference on Signal Processing and
  Communication (ICSC)}. IEEE, 2016, pp. 303--307.

\bibitem{barni2017eusipco}
Mauro Barni, Ehsan Nowroozi, and Benedetta Tondi,
\newblock ``Higher-order, adversary-aware, double {JPEG}-detection via selected
  training on attacked samples,''
\newblock in {\em 2017 25th European Signal Processing Conference (EUSIPCO)},
  Aug 2017, pp. 281--285.

\bibitem{ChenMFwithCNN}
J.~Chen, X.~Kang, Y.~Liu, and Z.~J. Wang,
\newblock ``Median filtering forensics based on convolutional neural
  networks,''
\newblock {\em IEEE Signal Processing Letters}, vol. 22, no. 11, pp.
  1849--1853, Nov 2015.

\bibitem{StammCNNuniv}
Belhassen Bayar and Matthew~C. Stamm,
\newblock ``A deep learning approach to universal image manipulation detection
  using a new convolutional layer,''
\newblock in {\em Proceedings of the 4th ACM Workshop on Information Hiding and
  Multimedia Security}, New York, NY, USA, 2016, IH\&\#38;MMSec '16, pp. 5--10,
  ACM.

\bibitem{barni2017aligned}
Mauro Barni, Luca Bondi, Nicol{\`o} Bonettini, Paolo Bestagini, Andrea
  Costanzo, Marco Maggini, Benedetta Tondi, and Stefano Tubaro,
\newblock ``Aligned and non-aligned double {JPEG} detection using convolutional
  neural networks,''
\newblock {\em Journal of Visual Communication and Image Representation}, vol.
  49, pp. 153--163, 2017.

\bibitem{simonyan2014very}
Karen Simonyan and Andrew Zisserman,
\newblock ``Very deep convolutional networks for large-scale image
  recognition,''
\newblock {\em arXiv preprint arXiv:1409.1556}, 2014.

\bibitem{Zuiderveld1994}
Karel Zuiderveld,
\newblock {\em Contrast Limited Adaptive Histogram Equalization},
\newblock Graphics Gems IV, Academic Press Professional, Inc., San Diego, CA,
  USA, 1994.

\bibitem{RAISE8K}
Duc-Tien Dang-Nguyen, Cecilia Pasquini, Valentina Conotter, and Giulia Boato,
\newblock ``Raise: A raw images dataset for digital image forensics,''
\newblock in {\em Proceedings of the 6th ACM Multimedia Systems Conference},
  New York, NY, USA, 2015, MMSys '15, pp. 219--224, ACM.

\bibitem{Keras}
Fran\c{c}ois Chollet et~al.,
\newblock ``Keras,'' \url{https://github.com/keras-team/keras}, 2015.

\end{thebibliography}

\end{document}